# Assessing the State of e-Readiness for Small and Medium Companies in Mexico: A Proposed Taxonomy and Adoption Model

Guillermo Rodriguez-Abitia
*ITESM Campus Estado de México*

Susana Vidrio
*Universidad de Colima*

Claudia Montiel-Sanchez
*ITESM Campus Estado de México*

# Assessing the State of e-Readiness for Small and Medium Companies in Mexico: A Proposed Taxonomy and Adoption Model

**Guillermo Rodríguez-Abitia**
ITESM Campus Estado de México
grdrz@itesm.mx

**Susana Vidrio**
Universidad de Colima
svidrio@ucol.mx

**Claudia Montiel-Sánchez**
ITESM Campus Estado de México
cmontiel@itesm.mx

**ABSTRACT**

Emerging economies frequently show a large component of their Gross Domestic Product (GDP) to be dependant on the economic activity of small and medium enterprises (SMEs). Nevertheless, e-business solutions are more likely designed for large companies. SMEs seem to follow a classical family-based management, used to traditional activities, rather than seeking new ways of adding value to their business strategy. Thus, a large portion of a nation's economy may be at disadvantage for competition. This paper aims at assessing the state of e-business readiness of Mexican SMEs based on already published e-business evolution models and by means of a survey research design. Data is being collected in three cities with differing sizes and infrastructure conditions. Statistical results are expected to be presented. A second part of this research aims at applying classical adoption models to suggest potential causal relationships, as well as more suitable recommendations for development.

**INTRODUCTION**

Firms face more challenging changes than ever in the way they compete in what is now a global and technology-flooded economy. New trade agreements emerge ever more often, and companies are no longer national or regional, but rather truly multinational. On the other hand, there is an increasing digital dependence to stay competitive, given the evolution that Information and Communications Technologies (ITCs) have experienced in their application for businesses, migrating from a focus on efficiency to one on effectiveness, and then moving on to innovation.

However, ITC adoption has not been undertaken evenly throughout the business universe. Most technology solutions have been adopted more quickly by those companies that have enough resources to start a long and expensive implementation process. However, the process is not equally easy for SMEs. Such firms possess not only far less resources, but also less awareness of the possibilities that ITCs offer for fostering competitiveness.

This shortcoming is exacerbated in Latin America, where there is a natural gap with developed nations where technology is produced and implemented quickly. However, the Latin American economies rest in great proportion on the activities of SMEs. According to the Official Diary of the (Mexican) Federation, firms are classified in four categories, depending on the number of people employed, and industry, as shown on table 1. Based on this classification, only 4% of the companies in Mexico are small and medium sized. Despite the fact that such number is so small in count, these firms employ 30% of the labor force in the country. Furthermore, these enterprises are responsible for producing 43% of the Gross Domestic Product of Mexico, as stated in the national economic census of 1999 (INEGI, 2000). Therefore, facing a loss of competitiveness due to a lack of ITC vision, and implementation to engage in electronic business strategies, could bring devastating consequences.

The state of e-business adoption and evolution has been studied narrowly in the case of Latin America, and this becomes more so for the case of SMEs. Palacios Lara and Kraemer (2003) made a very comprehensive review of the state of e-commerce in Mexico, providing information related to relevance factors such as economics, education, infrastructure, regional concentration of firms, and so on. It reviewed also the initiatives of e-government to foster e-commerce, and qualifies as "fair" the level of readiness for the development of B2B and B2C. There is a general conclusion that there is a great potential for e-commerce in Mexico.

|  | Industry | Trade | Services | GDP |
|---|---|---|---|---|
| Micro | 0-10 | 0-10 | 0-10 | 20% |
| Small | 11-50 | 11-30 | 11-50 | 21% |
| Medium | 51-250 | 31-100 | 51-100 | 22% |
| Large | > 251 | > 101 | > 101 | 37% |

**Table 1. Classification of Firms by Number of Employees in Mexico (Diario Oficial de la Federación, March 30$^{th}$ 1999)**

Grandon and Pearson (2004) undertook a similar study in Chile to observe SMEs' managers' perceptions about e-commerce adoption. Furthermore, they sought to identify and rank the factors that influenced adoption, combining constructs from prior models of research, including the theory of planned behavior (Ajzen, 1991), and the technology acceptance model (Davis, 1989), among others. However, their study is limited to SMEs in the Bio-Bio region of Chile, and they considered SMEs those with less than 500 employees. Based on the categorization shown in table 1, this study could include also micro firms, which are not the object of our analysis, having different characteristics. The study in Chile does not address the different levels of adoption as a result of the natural evolutionary process of e-readiness at the firm level of analysis. As a result of this study, perceived potential outcomes seem to have a great influence in adoption decisions.

**E-READINESS AND IT CAPABILITY**

Many models and guidelines for evaluating e-readiness have been developed, and proposed. The concept of e-readiness is generally associated to assessing the capacity of a country to undertake an e-commerce deployment effort. Such capacity at the firm level of analysis is generally known as IT capability.

In one study on e-commerce readiness undertaken by the World Information Technology and Services Alliance (WITSA, 2000), eight issues of concern are identified for policy makers to take into consideration so that adoption of e-commerce can be fostered: trust, technology, workforce, public policy, taxation, business processes, costs, and consumer attitudes. Additionally, the main barriers to e-commerce adoption are identified in the survey, as shown in figure 1.

WITSA and McConnell (2000) determined e-readiness of 42 countries, including the following factors in their assessment: connectivity, e-leadership, information security, human capital, and e-business climate. Four out these five factors are easily converted from a national unit of analysis to a firm unit of analysis, the only exception being e-business climate.

Some methodologies have been developed to measure the level of evolution where a company stands in a continuum for e-readiness or e-capability. Click & Mortar has identified four phases of evolution as follows: (1) brochureware / website, (2) internal systems /corporate portal, (3) e-commerce, and (4) e-business. Each phase is characterized by particular indicators.

Zhu and Kraemer (2002) developed a set of metrics to determine the level of e-commerce capability. These are comprised by four factors: information, transaction, customization, and supplier connection. Used properly, these metrics could be valuable to determine the level of evolution for a SME. We could expect to make minor adjustments for that purpose. Additionally, there are some issues that are also related to firm performance in that study, which pertain to IT infrastructure. Such issues are local area networks, personal computers, and IT intensity. This work controls for firm size, suggesting that there is a potential for significant differences based on such factor. However, no conclusions can be drawn from that research work given that the control consisted on limiting their sample to Fortune 1000 companies, which by nature are only large organizations.

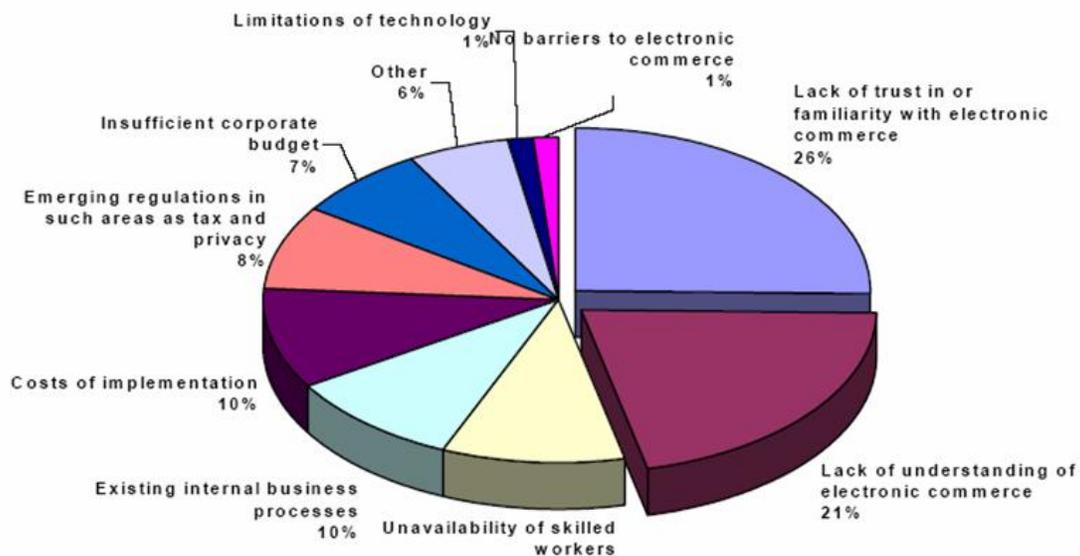

**Figure 1. Barriers faced by e-commerce (WITSA, 2000)**

**TECHNOLOGY ADOPTION FACTORS**

Many models have been developed, and proposed for explaining adoption behavior of IT in organizations. Some of the most classical are the Technology Acceptance Model (TAM) (Davis, 1989), the Theory of Reasoned Action (Fishbein and Azjen, 1975), the Theory of Planned Behavior (Azjen, 1991), and the Motivation Model (Davis, Bagozzi and Warshaw, 1992), among others. Individual factors that affect adoption have also been widely studied, being the most representative work perhaps the one of Rogers (1983) who analyzed prior models and identified five common factors of adoption: relative advantage, compatibility, observability, complexity, and trialability.

Harrison, Mykytyn and Riemenschneider (1997) assert that IT research in small business was scarce and mostly exploratory. Therefore, they went on to study IT adoption decisions in small firms, by combining the theory of planned behavior (Azjen, 1991), with the Technology Acceptance Model (Davis, 1989). They considered that such combination would fit better the small business context, obtaining a more powerful model, and more explanatory power. They found adoption decisions to be highly influenced by consequence perceptions, social influence, and resources to overcoming obstacles. This latter effect is in great concordance with the perception results obtained by Grandon and Pearson (2004).

More recently, Gefen, Karahanna, and Straub (2003) made an effort to combine perspectives in order to obtain a more comprehensive model to explain IT adoption. However, this time adoption was not the only action undertaken, but rather the use of potential capabilities. They incorporated Trust to the TAM perspective to study on-line shopping behavior.

Venkatesh, Morris, Davis, and Davis (2003) performed a review of eight previous models of adoption to obtain a unified view of adoption theory. They established correspondence among the constructs of the models, where applicable, and consolidated all the constructs obtained in one model which was empirically validated, both in cross-sectional and longitudinal studies. They named the resulting model as "Unified Theory of Acceptance and Use of Technology" (UTAUT). The model is comprised by four constructs: performance expectancy, effort expectancy, social influence, and facilitating conditions. There are also four moderators of the relationships: gender, age, experience and voluntariness of use. Figure 2 illustrates the model.

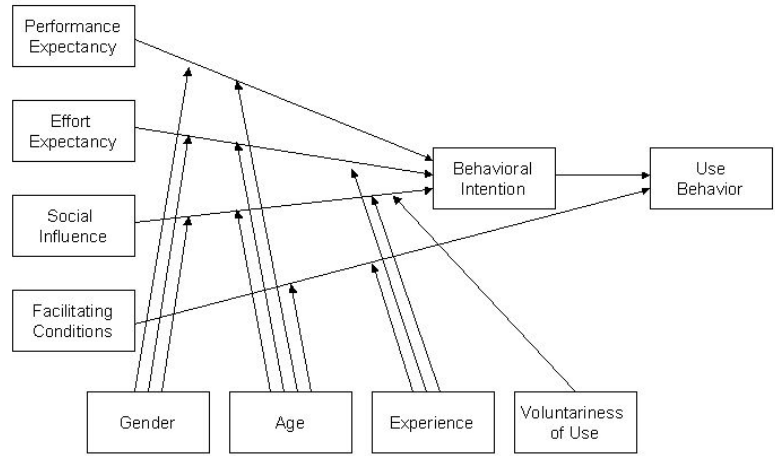

**Figure 2. Unified Theory of Acceptance and Use of Technology (Venkatesh, et. al., 2003)**

**RESEARCH MODEL**

It is reasonable to think that environmental conditions, such as existing local infrastructure, as well as the availability of qualified workforce in a community or region may influence greatly the effect on e-readiness. Although facilitating conditions are included in the UTAUT model proposed by Venkatesh, et.al. (2003), those conditions are related to those provided within the enterprise to foster the acceptance of the technology in question, not to factors of local infrastructure. This is of particular importance in Latin America, where most infrastructure conditions are present in large urban areas, but they are not equally distributed to smaller communities.

Our research strategy aims at attaining three main objectives:

1. Applying a taxonomy to classify the level of e-readiness at the firm level, tailored specifically to SMEs in Mexico
2. Understanding the factors that foster or inhibit e-commerce adoption in SMEs in Mexico.
3. Identifying potential differences in the adoption of e-commerce in urban communities of different sizes.

The general research strategy is illustrated in figure 3.

The instrument for evaluating the level of evolution of e-readiness were, based on the metrics for e-commerce readiness developed by Zhu and Kraemer (2002) has four factors that will be considered for discrimination: information, transaction, customization, and supplier connection. The instrument for understanding the acceptance factors was based on the one used by Venkatesh, et. al. (2003) to test the UTAUT model. Other metrics were included in the instrument, adapted from Malhotra and Galletta, 1999. Two pilot studies have been conducted to assess the validity and reliability of the instrument, and the appropriate adjustments have been performed. Power analysis was undertaken to determine sample size.

In order to observe possible differences based on regional infrastructure and development conditions, a stratified random sample has been drawn from localities that constitute the metropolitan areas of three cities with differing sizes and conditions, and data collection is in progress.

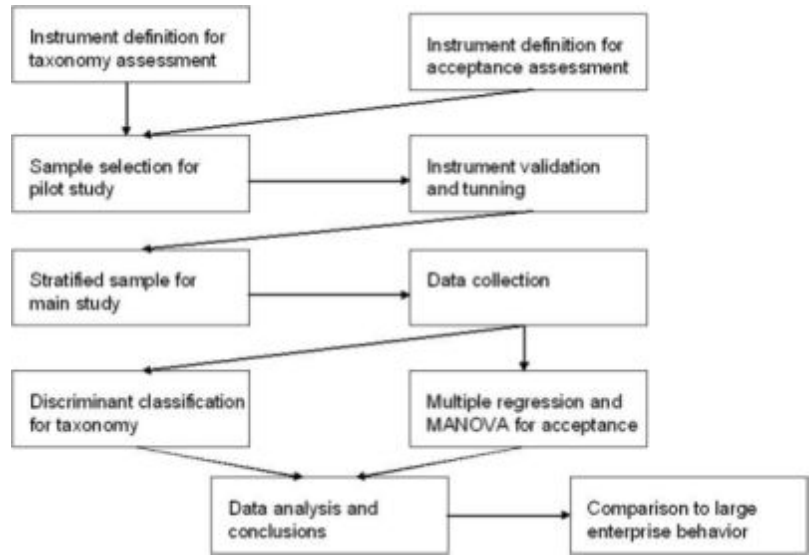

**Figure 3. Research Strategy**

**CONCLUSION**

To foster regional development, it is very important to undertake research projects that not only address relevant issues of IT implementation, but also take into consideration contextual characteristics. Often, research results obtained for specific environments are not easily generalizable. Most research is geared towards large organizations in developed nations where optimal conditions for IT implementation can be found. Conducting studies that target Small and Medium Enterprises in Latin American countries is of a rather crucial importance. We believe our work will yield very interesting conclusions that will allow the establishment of concrete actions to foster competitiveness based on IT in Mexico. It may be expected to find that the state of e-readiness of Mexican SMEs is far less developed than that found in developed nations. The revised theory states the great impact that perceptual issues have on IT adoption decisions. It may be expected that Latin American expectations of the potential benefit of e-commerce solutions may differ greatly from those in developed nations, having a dampening effect on e-commerce adoption. Other dampening effects may include infrastructure and perception conditions that vary depending on the community under study.

**REFERENCES**

1. Ajzen, I. (1991) The Theory of Planned Behavior, *Organizational Behavior & Human Decision Processes*, 50, 2, 179-211.
2. Click & Mortar (2004) *e-Business Evolution – Phases* available at:
   http://www.click&mortar.com
3. Davis, F. D. (1989) Perceived Usefulness, Perceived Ease of Use, and User Acceptance of Information Technology, *MIS Quarterly*, 13, 3, 319-339.
4. Davis, F. D., Bagozzi, R. P., and Warshaw, P. R. (1992) Extrinsic and Intrinsic Motivation to Use Computers in the Workplace, *Journal of Applied Social Psychology,* 22, 14, 1111-1132.
5. Diario Oficial de la Federación (1999) Gobierno de la República, México, D.F., March 30th.
6. Fishbein, M., and Azjen, I. (1975) Belief, Attitude, Intention and Behavior: An Introduction to Theory and Research, Addison-Wesley.
7. Gefen, D., Karahanna, E., and Straub, D. W. (2003) Trust and TAM in Online Shopping: An Integrated Model, *MIS Quarterly,* 27, 1, 51-90.
8. Grandon, E., and Pearson, J. M. (2004) E-commerce Adoption: Perceptions of Managers / Owners of Small and Medium Sized Firms in Chile, *Communications of the Association for Information Systems*, 13, 81-102.

9. Harrison, D. A., Mykytyn, P. P., and Riemenschneider, C. K. (1997) Executive Decisions About Adoption of Information Technology in Small Business: Theory and Empirical Tests, *Information Systems Research*, 8, 2, 171-195.
10. INEGI (2000), *Censos Económicos 1999*, Instituto Nacional de Estadística, Geografía e Informática, Aguascalientes, Ags.
11. Malhotra, Y. and Galletta, D. (1999) Extending the Technology Acceptance Model to Account for Social Influence: Theoretical Bases and Empirical Validation, *Proceedings of the 32$^{nd}$ Hawaii Conference of System Sciences*, Maui, Hawaii.
12. McConnel & WITSA (2000) *Risk e-Business: Seizing the Opportunity of Global e-Business*, available at: http://www.mcconnellinternational.com/ereadiness/ereadiness.pdf
13. Palacios Lara, J. J., and Kraemer, K. L. (2003) Globalization and E-Commerce IV: Environment and Policy in Mexico, *Communications of the Association for Information Systems*, 11, 129-185.
14. Rogers, E. M. (1983) Diffusion of Innovations, Free Press, New York, N.Y.
15. Venkatesh, V., Morris, M. G., Davis, G. B., and Davis, F. D., (2003) User Acceptance of Information Technology: Toward A Unified View, *MIS Quarterly*, 27, 3, 425-478.
16. WITSA (2000) *International Survey of e-Commerce*, available at: http://www.witsa.org/papers/EComSurv.pdf
17. Zhu, K., and Kraemer, K. L., (2002) e-Commerce Metrics for Net-Enhanced Organizations: Assessing the Value of e-Commerce to Firm Performance in the Manufacturing Sector, *Information Systems Research*, 13, 3, 275-295.